\documentclass[useAMS,usenatbib,usegraphicx]{mn2e}
\usepackage{booktabs} 
\usepackage[dvips]{graphicx}
\usepackage[usenames,dvipsnames]{color}
\usepackage[pdftex]{hyperref}
\usepackage{amssymb} 
\usepackage{amsmath}
\usepackage{times}
\title{On the Reported Death of the Macho Era }
\date{   }
\author[Quinn et al.]{D.P. Quinn$^{1}$, M. I. Wilkinson$^{2}$, M. J. Irwin$^{1}$,
J. Marshall$^{3}$, A. Koch$^{2}$,V. Belokurov$^{1}$\\
$^{1}$ Institute of Astronomy, University of Cambridge, Madingley Road,
Cambridge CB3 0HA, UK \\
$^{2}$ Dept.\ of Physics and Astronomy, University of Leicester,
University Road, Leicester LE1 7RH, UK \\
$^{3}$ Dept.\ of Physics, Texas A\&M University, 4242 TAMU, College Station, TX 77843-4242, USA\\
}


\begin{document}
\maketitle
\begin{abstract}
We present radial velocity measurements of four wide halo binary
candidates from the sample in Chaname \& Gould (2004; CG04) which, to
date, is the only sample containing a large number of such
candidates. The four candidates that we have observed have projected
separations $>0.1$ pc, and include the two widest binaries from the
sample, with separations of 0.45 and 1.1 pc. We confirm that
three of the four CG04 candidates are genuine, including the one with
the largest separation.  The fourth candidate, however, is spurious at
the 5-sigma level.  In the light of these measurements we re-examine
the implications for MACHO models of the Galactic halo. Our analysis
casts doubt on what MACHO constraints can be drawn from the existing
sample of wide halo binaries.
\end{abstract}

\section{Introduction}

Although convincing evidence for the existence of dark matter has been
around for over 40 years, its nature remains a mystery. If MAssive
Compact Halo Objects (MACHOs) constitute a significant fraction of the
dark matter budget, then a combination of observational and
theoretical arguments constrain the properties of viable MACHO
candidates to well-defined regions of parameter space.  Microlensing
experiments
\citep[eg][]{lukasz,Tisser} have, for instance, ruled out MACHOs with
masses in the range $10^{-7}-30$ M$_{\odot}$ as major constituents of
the Milky Way's dark matter halo, thereby excluding dark matter
candidates such as halo brown dwarfs or solar-mass black holes. In
addition to microlensing, constraints from a number of indirect
arguments such as the observed velocity dispersion in the disk
\citep{lacey}, evaporation of low mass gas clumps (''snowballs'')
\citep{rujula}, further reduce the parameter space available to
baryonic Galactic MACHOs to $\approx 30-10^{6}$M$_{\odot}$.

A recent analysis of the distribution of wide halo binaries in
\citeauthor{Yoo}~(\citeyear{Yoo}; hereafter Yoo04), failed to detect 
a clear signature of the disrupting effect of MACHOs on the widest,
and hence most weakly bound, binaries. Consequently, the study
appeared almost to close the door entirely on the remaining region of
viable MACHO parameter space, leaving only a small window between 30
$M_{\odot}$ and 43 $M_{\odot}$.  A look, though, at Fig.5 in Yoo04
suggests that their results depend critically on the validity of the
two widest binaries in the observed wide halo binary sample from
\citeauthor{Chan}~(\citeyear{Chan}; hereafter CG04).

In this letter, we present radial velocity measurements of the stars
in each of these candidate binaries along with two other
large-separation halo binary candidates from CG04. Our radial
velocities imply that three of these candidates are genuine binaries
and we thus demonstrate directly that halo binaries with projected
separations of $\approx$ 1 pc exist.  However, our measurements also
reveal that the second widest binary in CG04 is actually a spurious
interloper. We update the constraints on MACHOs arising from these
measurements. The removal of the spurious pair from the analysis eases
the constraints significantly with the upper limit on MACHO mass
increasing by an order of magnitude. In addition, the Galactic orbit
we obtain for the widest binary raises questions on the validity of
using this object in the analysis carried out by Yoo04. Its omission
would re-open the region of parameter space closed in Yoo04.
Furthermore, we also point out that, if the initial logarithmic slope
of the binary separation function is set to -1, a choice with some
theoretical foundation, then an un-evolved distribution is ruled out
by the observations.

The outline of this letter is as follows. In Section~\ref{sec:data} we
present our spectroscopic data for the candidate binaries, and in
Section~\ref{sec:vels} we derive radial velocities for the pairs and
use these to determine which systems are genuine binaries. In
Section~\ref{sec:macho}, we re-visit the constraints on the MACHO
content of the Milky Way halo based on our new
data. Section~\ref{sec:conc} summarises our conclusions.

\section{Radial Velocities of Wide Halo Binary Candidates in CG04 Sample}
\label{sec:data}
The search for wide halo binaries is still in its infancy as a number
of difficult observational challenges need to be overcome.  First,
halo stars are rare, constituting less than 0.2$\%$ of local stars
\citep{helmi_review}. Second is the problem of distinguishing wide
binary stars in samples of halo stars from mere chance associations.
To date there has been one keynote study (CG04), which detected a
large number of local (sample median distance is 240 pc),
high probability candidate wide halo binaries (namely
116).  The angular separation function of these binaries followed a
power law distribution out to angular separations corresponding to
$\approx$1 pc which was the detection limit of their survey.

The candidate binaries in the CG04 sample were chosen from the revised
New Luyten Two-Tenths Catalog (NLTT) of high proper motion stars,
\citep{rNLTT,rNLTT2}.  Candidate halo pairs were required to satisfy
proper motion consistency tests and to lie along what are essentially
isochrones in the halo region of the reduced proper motion (RPM)
diagram.  As the angular separation, $\Delta \theta$, between the
members in a candidate binary increases, the probability that the
candidate pair is merely a random association increases roughly
in proportion to $(\Delta \theta)^2$. CG04 argue that their
halo binary sample is unlikely to be contaminated out to $\Delta
\theta=900
\arcsec$.  While the CG04 candidate binaries have angular separations
smaller than this value, their importance for placing restrictions on
viable MACHO candidates for dark matter demands that they be subject
to further tests, in particular the candidate binaries with the widest
angular separations which run the largest risk of misidentification.

One useful test to further explore the nature of these objects is to
measure their radial velocities: both members of the candidate
binaries should have essentially identical velocities on account of
the long period of wide binaries. For example, the magnitude of the
relative velocity for a binary of mass $m$ with separation $a$
(assuming circular orbits) is given by
\begin{equation}\label{bin_rot}
v=0.2\textrm{km/s}\sqrt{\frac{m}{M_{\odot}}}\left(\frac{a}{0.1\textrm{pc}}\right
)^{-1/2}.
\end{equation}
Combined with the other information, radial velocities also enable the
Galactic orbital properties of these objects to be studied.

A search in SIMBAD of the 11 objects with separations $>100\arcsec$
reveals radial velocity data for each member of the candidate binary
NLTT 39456/39457. The radial velocities agree at the 3$\sigma$ level
and the members also have consistent parallaxes - thus, this candidate
is almost certainly a binary.  The orbit of this object has been
analysed in detail in
\cite{Allen} and provides strong evidence that wide 
halo binaries with separations out to at least $\approx 0.05$pc 
exist.

In order to probe further the other candidate wide binaries we set out
to obtain spectra for a representative sample of the CG04 sample with
angular separations $>100\arcsec$ that includes the 2 widest halo
binary candidates. We obtained spectra for four candidate pairs,
listed in Table~\ref{CGtable} and referenced by their identifiers in
the NLTT catalog. Using the colour-absolute magnitude relation given
in CG04 the 4 candidates NLTT 1715/1727, 10536/10548, 15501/15509 and
16394/16407 have projected separations of 0.45, 0.2, 0.14 and 1.1 pc,
respectively.  The observations of NLTT 1715/1727, 10536/10548,
15501/15509 were carried out on the William Herschel Telescope (WHT,
La Palma) as part of the Isaac Newton Group (ING) service observing
program. NLTT16394/16407 was observed with the Magellan
telescope.

\subsection{Data Reduction } 
\subsubsection{WHT spectra}
The pairs NLTT 10536/10548 and NLTT 15501/15509 were observed with the
single-slit, Intermediate dispersion Spectrograph and Imaging System
(ISIS) mounted on the WHT, on November 27, 2007 while the pair NLTT
1715/1727 was observed on July 23, 2008.  The red and blue arm of the
spectrograph were used to provide a resolution of about 1\AA\, and
spectral coverage over $3587-5412$\AA\, and $7587-8812$\AA\,, the
latter to cover the Ca II triplet.  (The blue region suffered from
vignetting and was ignored in the analysis).

\begin{figure}
\includegraphics[width=0.48\textwidth]{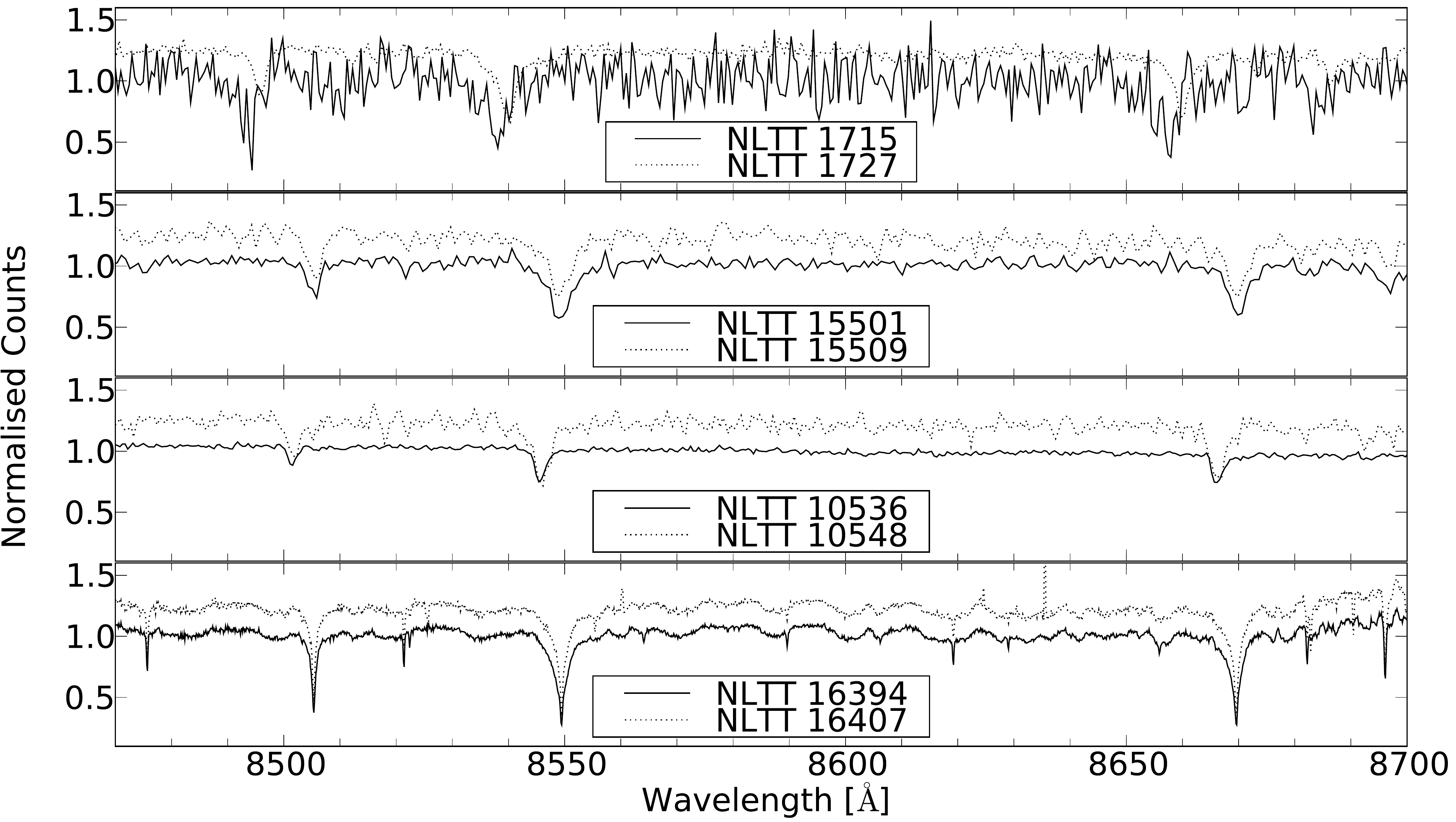}
\caption[]{\label{wide_spec} Spectra for the four
candidate binary systems we have observed. The top three panels show
data from our WHT observations, while the bottom panel shows the data
for the pair NLTT 16394/16407 taken with the Magellan telescope. In
each panel, the spectra are centred on the Ca II triplet region. The
spectra for each member of a candidate binary are plotted
together. The second component is shifted slightly upwards for
clarity.}
\end{figure}

Two exposures for each science image were taken.  S/N at $8600$\AA\,
varied from $\approx$ 20 for the fainter objects to $\approx$ 60 for
the bright objects. The spectra were reduced following standard steps
using IRAF.  Each science image was bias-subtracted and
flat-fielded. The iraf task {\tt apall} was used to extract the
sky-subtracted image along with a spectrum of the sky taken from
regions of the slit adjacent to, but not dominated by, the light of
the target.  Typically 5-6 un-blended sky lines spanning the
wavelength range of the Ca II triplet were identified with the help of
the sky-line catalog in \cite{skylines} and used to define a
wavelength solution (using the {\tt identify} and {\tt reidentify}
tasks in IRAF).  A Chebyshev function of order 2 was used in this step
to map from pixels to wavelength; the typical error on the wavelength
solution was about $0.2$\AA\, i.e., about 7 km/s at the wavelengths of 
the triplet. After applying the
appropriate wavelength solution to each science image, the wavelength
shifts in the Ca II triplet lines were determined. This step was carried
out interactively within the IRAF tool {\tt splot}; the prominent
triplet lines were identified and the {\tt d} function in splot was
used to find the center of each of the lines (by fitting a Gaussian to
the line).  An average of the velocity shifts of the 3 lines for each
of the 2 exposures was used to determine the radial velocity; the standard 
deviation of the measurements was used to define the velocity error. We
convert these to heliocentric velocities using IRAF task {\tt
rvcorrect}.  The results are listed in Table~\ref{CGtable} and sample
spectra are given in the top 3 panels of Figure~\ref{wide_spec}.

\begin{table*}
\small
\begin{tabular*}{\textwidth}{@{\extracolsep{\fill}}    c c  c  c c c c c c c c c c c c c  c c c }
\toprule
NLTT ID& &1715 & 1727& 10536 & 10548&15501&15509& 16394& 16407\\
\cmidrule(r){1-2}
\cmidrule(r){3-4}
\cmidrule(r){5-6}
\cmidrule(r){7-8}
\cmidrule(r){9-10}
Position&$\alpha$&   7.98335&   8.03731& 49.62049&  49.67233&85.91593&  85.97542&94.91613&95.11188  \\ 
(J2000)& $\delta$ & -10.71683&  -10.83106&-7.14044&  -7.13639& 49.38367& 49.37782&-30.70087&-30.60432  \vspace{0.0cm}\\
Proper motion &$\mu_{\alpha}\textrm{cos}(\delta)$& -0.034& -0.035&0.171& 0.164&0.081&0.081&0.328&0.325  \\
(arcsec/yr)  &$\mu_{\textrm{Dec}}$&-0.390&-0.383&-0.353&-0.347&-0.176&-0.182&-0.172&-0.163\vspace{0.0cm}\\
Magnitude& V& 17.6 & 16.1 &11.22&15.8&17.2&17.6&12.22&15.33\vspace{0.0cm}\\ 
Colour&V-J&2.50&2.46&0.98&2.29&2.82&2.97&0.98&2.19\vspace{0.0cm}\\
Distance& pc& 209&209&219&219&210&210&348&348\vspace{0.1cm}\\
Pair Separation& arcsec& 453.4&453.4&185.7&185.7&141.0&141.0&698.5&698.5\vspace{0.0cm}\\
Radial Velocity&km/s&-123.2$\pm$13.7&-45.6$\pm$9.1& 121.6$\pm$6.8 &122.6$\pm$7.2&  262.3$\pm$10.5 & 265.2$\pm7$&268.2$\pm1.7$ &268.3$\pm1.4$\\
P(B$|\Delta v_r$) & & & \hspace{-2cm} 0.001  & &\hspace{-2cm} 0.993& &\hspace{-2cm} 0.991  & &\hspace{-2cm} 0.999 

\end{tabular*}
\caption{Table listing, for convenience, the properties of the four
candidate wide binaries, taken from the compilation given in
CG04. Also included are the measured heliocentric radial velocities,
an estimate of the distance to the putative binary based on applying the CG04
photoparallax relation to the brightest member of each candidate, and
the probability P(B$|\Delta v_r$) that the pair is a genuine binary
(equation~\protect\ref{prob_real}).}
\label{CGtable}
\end{table*}

\subsubsection{Magellan Spectra}

Observations of the NLTT 16394/16407 pair were taken on November 13,
2008 with the Magellan Inamori Kyocera Echelle (MIKE) spectrograph at
the 6.5-m Magellan2/Clay Telescope.  We used a slit width of
0.7$\arcsec$ and a binning of 2$\times$2 CCD pixels in the spatial and
spectral dimensions, from which we obtain a spectral resolving power
of R$\sim$30,000.  As for the WHT data, we employed the red and blue
CCDs of the instrument, yielding a total wavelength coverage of
3340--9150\AA.  While NLTT\,16394 was exposed for 2$\times$300\,s, we
acquired 2$\times$900\,s exposures of the fainter NLTT\,16407.  The
pipeline reduction package of
\citeauthor{kelson00}~(\citeyear{kelson00,kelson03}) was used to
reduce and extract the spectra from the raw data. Wavelength
calibration was carried out via built-in Th-Ar lamp exposures, taken
between each pair of science exposures.  As a result, we reach
signal-to-noise (S/N) ratios of 115 (55) per pixel at
6500\AA~(4000\AA) for NLTT\,16394 and 60 (10) per pixel for
NLTT\,16407.  We determined the radial velocity again by averaging the
Doppler shift of the CaII triplet lines.

\subsection{Diagnosis from Radial Velocities}
\label{sec:vels}
If the objects are genuine binaries with separation $a>0.1$ pc, the
measurement errors will dominate the differences between the measured
radial velocities of the stars, $\Delta v_{r}$, so the probability of
measuring a particular value of $\Delta v_{r}$ for a true binary,
P($\Delta v_{r}|$B), will be a Gaussian with mean zero and dispersion
given by the radial velocity measurement error. If we assume the
objects are halo stars (i.e.  the CG04 classification is correct) and,
furthermore, that the stellar halo relative velocity distribution is a
Gaussian with dispersion $\sqrt{2}\times116$\,km/s \citep[i.e. we use the
root mean square dispersion from the observed triaxial dispersion
tensor in][]{chiba}, then the probability of measuring $\Delta v_{r}$
if it is a spurious binary, P($\Delta v_{r}|$Not B), will also be
Gaussian with dispersion dominated by the relative halo radial
velocity dispersion. We can write the probability that the binary is
real given the new data as
\begin{equation}\label{prob_real} 
P(\textrm{B}|\Delta v_{r})=\frac{P(\Delta v_{r}|\textrm{B})}{P(\Delta v_{r}|\textrm{B})P(\textrm{B})+P(\Delta v_{r}|\textrm{Not B})P(\textrm{Not B})}P(\textrm{B})
\end{equation}
where P(\textrm{B}) is the prior probability that the object is a
binary. Given that CG04 estimated that about one false
detection would be expected in their halo sample, we take
P(\textrm{B}) to be 10/11.  Using Equation~\ref{prob_real} we list
the probabilities that the objects are binaries in
Table~\ref{CGtable}.

The results show that we have at high probability confirmed the binary
nature of three of the pairs in the CG04 sample.  Even if we reduce
$P(\textrm{B})$ to the extremely conservative value of 0.5, the
$P(\textrm{B}|\Delta v_{r})$ is still greater than 0.9 for the 3
objects.  NLTT 1715/1727, the second widest pair, turns out to have
inconsistent radial velocities (at the $5\sigma$ level). This was
flagged by CG04 as potentially spurious on the basis of the pair's
position in the RPM diagram but was nonetheless used in the analysis
of \cite{Yoo}.  This result underscores the importance of conducting
follow-up observations of the CG04 sample. However, we
emphasise that our results are entirely consistent with the estimation
of one false detection in CG04.

Most importantly, the observation of the widest halo binary candidate
in CG04, with projected separation 1.1 pc, reveals that it is in fact
a true halo binary. Its Galactic orbit, discussed below, is also
consistent with this interpretation. This result provides
strong evidence that wide halo binaries with $a\gtrsim 1$ pc can
exist. In the next section we consider implications of our
measurements for the constraints on the MACHO content of the Milky Way
halo. An investigation of the origin of binaries with such wide
separations is deferred to a future paper (Quinn et al., in prep.).

\section{Re-examining Dark Matter Constraints from the CG04 Sample}
\label{sec:macho}
Weakly-bound wide binaries are vulnerable to disruption from
encounters with massive compact objects.  Depending on the properties
of the perturbers and the fragility of the binary star, encounters
fall into two regimes: a diffusive regime in which the typical change
in binding energy of the binary induced by an encounter is small in
magnitude relative to its binding energy, and a catastrophic regime in
which the energy change from the closest encounter can disrupt the
binary.  Expressed in terms of fiducial values of potential MACHO
parameters, the disruption timescales for a solar mass binary
with separation of order 0.1 pc in each of these regimes,
$t_{\textrm{diff}}$ and $t_{\textrm{cat}}$, are~\citep[see Equations
8.65a and b of][]{BT}
\begin{equation}
t_{\textrm{diff}}\approx \frac{v_{p}}{\textrm{200
km/s}}\frac{100M_{\odot}}{M_{p}}\frac{0.01M_{\odot}\textrm{pc}^{-3}}{\rho_{p}}\frac{0.1\textrm{pc}}{a}\textrm{
Gyr}
\end{equation}
and 
\begin{equation}
t_{\textrm{cat}}\approx 3\left(\frac{0.01M_{\odot}\textrm{pc}^{-3}}{\rho_{p}}\right)\left(\frac{0.1\textrm{pc}}{a}\right)^{3/2}\textrm{ Gyr}
\end{equation}
where $v_{p}$ is the relative velocity dispersion between MACHO
perturbers and binaries, $M_{p}$ is the mass of the perturber and
$\rho_{p}$ is the density of perturbers.  The perturber mass,
$m_{\textrm{crit}}$, which marks the transition between the two
regimes is
\begin{equation}
m_{crit}\approx 30M_{\odot}\frac{v_{p}}{200 \textrm{km/s}}\left(\frac{a}{0.1\textrm{pc}}\right)^{1/2}
\end{equation}
Evaluating the timescales at the fiducial values we find that the disruption timescale for binaries with 
separation of 0.1 pc or more is shorter than 10 Gyr if the perturber 
mass is greater than about 10 $M_{\odot}$. We can thus expect to see 
a signature on the binary separation function at these scales if a 
population of MACHOs with masses above this threshold exists. 

The study by Yoo04 found no strong signature in the sample of CG04 and
consequently was able to rule out MACHOs with masses above 43
$M_{\odot}$.  In this section we reassess that analysis and the
constraints on MACHOs from the CG04 dataset in the light of our
measurements.  We focus on exploring how the MACHO mass and the
fraction of the halo composed of MACHOs affect the observed binary
separation function.

As our treatment follows closely the analysis in Yoo04 we only briefly
summarize our analysis procedure. We assume the original binary
separation function is a power law of the form $f(a)\propto
a^{-\alpha}$, with $\alpha$ to be determined from the observations.
The final binary separation function is determined by simulating
encounters between perturbers and binaries, using the impulse
approximation to work out the change in binding energy induced by each
encounter.  We assume the total mass of the binary is 1 $M_{\odot}$,
take the 1-D relative velocity dispersion between MACHO
perturbers and the binaries to be 200 km/s and assume that the
binaries are subject to encounters for a period of 10 Gyr. These
assumptions are consistent with the choices in Yoo04.
We also assume that the binaries move through a constant halo density
which we take to be the local dark matter density assumed to be
$0.01M_{\odot}\textrm{pc}^{-3}$, the value in Milky Way Mass Model 1
of \cite{DehnenMW} which was one of their best-fit models.  The value
of the local dark matter density and velocity dispersion are within
3$\sigma$ of the values found in a recent Bayesian analysis of Milky
Way models by \cite{Wid08}, which indeed also re-affirmed the Milky
Way models of
\cite{DehnenMW}.  As we can see from the timescales above, the choice
of background perturber density is crucial; in the catastrophic regime
this parameter sets the disruption timescale for a given total binary
mass.

We adopt the procedure given in Yoo04 to generate the model wide
binary angular separation function, P(log$(\Delta
\theta)|M_{p},\rho_{p},\alpha$) (i.e. depending on perturber mass,
density of perturbers and power law exponent for the initial binary
distribution) from the model separation function; this involves
convolving the model binary separation function with the distance
distribution of the observed wide binaries.

We choose the normalisation so that sum of $\textrm{P}(\textrm{log
}\Delta \theta)\Delta(\textrm{log }\Delta \theta)$ over 24 angular
separation bins, equally spaced logarithmically between 3.5$\arcsec$
and 900$\arcsec$ (corresponding to the interval in angular separation
over which CG04 claim to have a clean and complete sample of wide
halo binaries), equals the number of observed binaries in this range.
\begin{figure}
\includegraphics[width=0.47\textwidth]{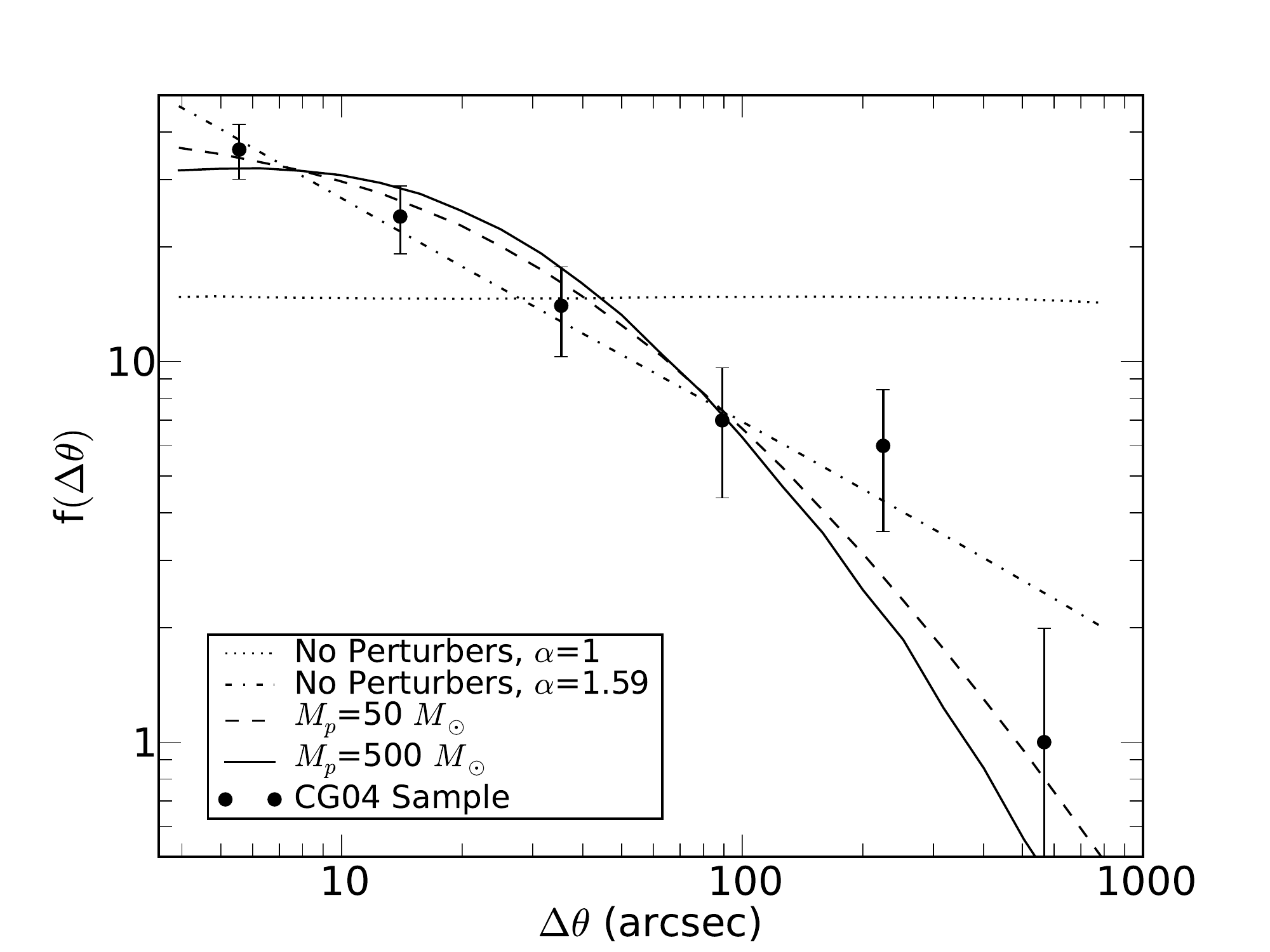}
\caption[]{\label{fang_comp}A comparison of the predicted observable 
angular separation function for a number of models. See text for
details.  The observed distribution, from the CG04 homogeneous sample
less the spurious pair, is also shown with associated Poisson
errors.}
\end{figure}
The log likelihood, $\log L$, of the model parameters given the data 
is then 
\begin{equation}
\textrm{log }L=\sum_{k=1}^{24}n_{k}\textrm{log }\textrm{P}(\textrm{log}\Delta\theta_{k})
\end{equation}
with $n_{k}$ the number of halo binaries from CGO4 in the bin centered on 
$\textrm{log} \Delta\theta_{k}$.

We explore the combined constraints on the mass and the density of a
putative perturber population by maximising the likelihood over
$\alpha$.  In figure (\ref{fang_comp}) we plot the model predictions
for the angular separation distribution assuming MACHO halo mass
fraction of unity and values of 50 and 500 for $M_{p}$ for which we
find $\alpha$ to be 1.06 and 0.80, respectively. We also show
a model with no perturbers finding in this case 1.59 for
$\alpha$. The plot clearly shows that the main question at stake is
whether the data favour a flat power law in the inner regions that
becomes steeper through the action of perturbers or essentially a
featureless power law for the case without perturbers.

Figure~\ref{fang_comp} also shows the unevolved case for
$\alpha=1$. This is the value favoured by observations of disk
binaries with separation greater than 100 AU and could be the outcome
of energy relaxation processes in the formation of wide binaries
\citep{lepine}. The data are clearly inconsistent with an
unevolved distribution with an initial value of $\alpha=1$.

\subsection{Updating the constraints on MACHOs}
The analysis of the CG04 sample by Yoo04 favored a model with no
perturbers. In conjunction with microlensing experiments their results
rule out most of the available parameter space for halo models which
have a significant contribution from baryonic MACHOs. At the 95$\%$
percent confidence level only a small window,
$30M_{\odot}<M_{p}<43M_{\odot}$ was left open for haloes composed
entirely of MACHOs.

\begin{figure}
\includegraphics[width=0.44\textwidth]{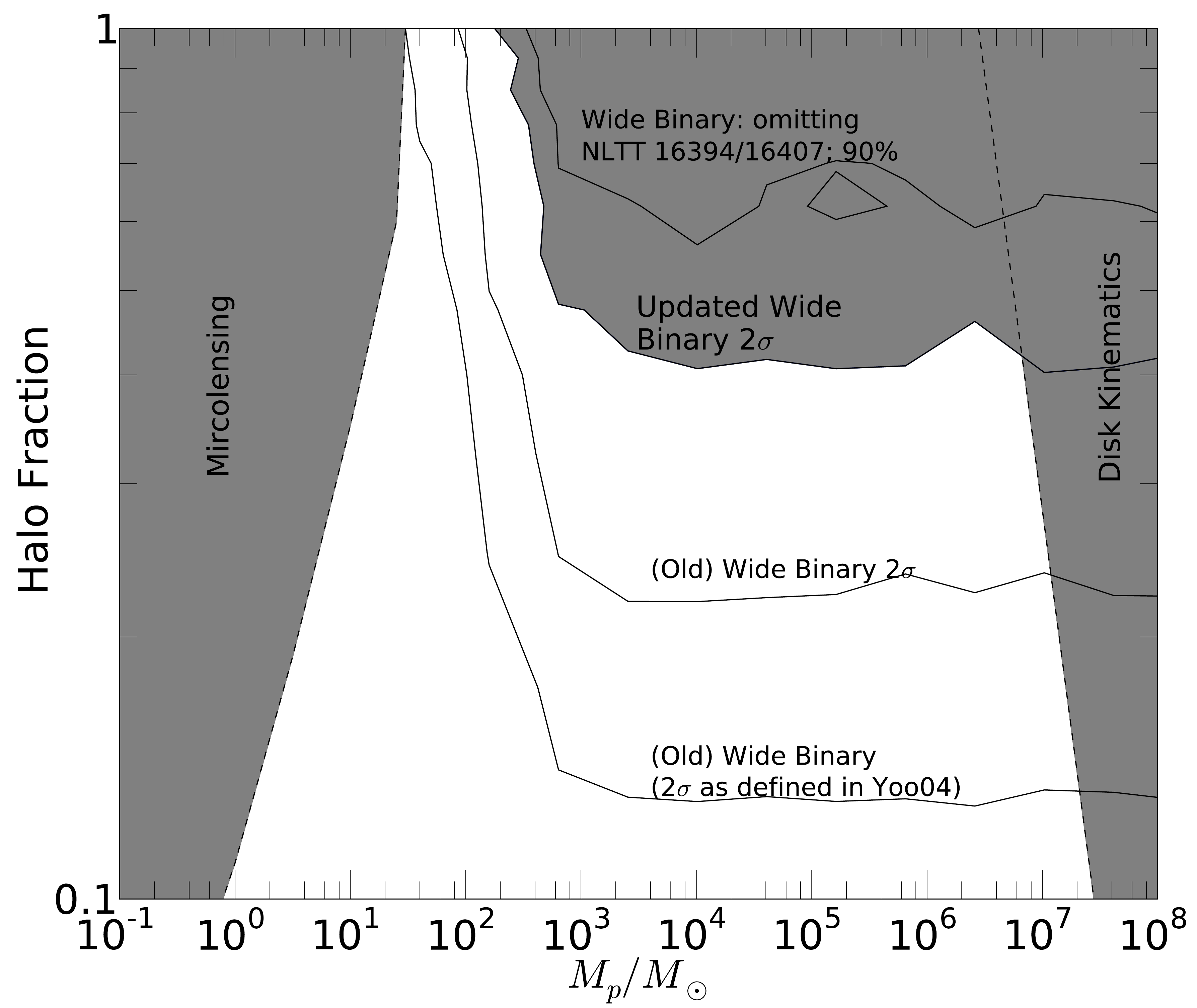}
\caption[]{\label{macho_conf} Confidence regions for the 
MACHO mass $M_{\rm p}$ and MACHO halo fraction.  For comparison to
past work we show the $2\sigma$ joint confidence levels as defined in
Yoo04 and using the standard definition, respectively, when the whole
CG04 homogeneous sample is included.  We also show the updated
$2\sigma$ confidence levels omitting the spurious candidate
binary. The omission of this object eases the constraints on MACHOs;
the window increases to $\approx 30-500$ M$_{\odot}$. In addition, the
effect on the constraints of omitting the widest binary in CG04 is
shown at the 90\% confidence level: the constraints at the $2\sigma$
level vanish.  The regions of parameter space shaded in grey are ruled
out at the $2\sigma$ level by binaries and microlensing data -- an
upper limit on the MACHO mass and halo fraction from disk kinematics
is also shown.  We stress that the constraints from the binaries are
based on the assumption that the time-averaged dark matter density
experienced by each binary is the local halo density at the position
of the Sun -- the actual Galactic orbits of the confirmed wide
binaries suggest much lower time-averaged dark matter densities. See
text for a detailed discussion. }
\end{figure}

In Figure~\ref{macho_conf} we replot the constraints on MACHO 
mass and halo fraction from the complete CG04 homogeneous 
sample. We show contours for the joint 95$\%$ confidence levels 
using the definition adopted in Yoo04 (which strictly 
only holds for 1d confidence levels) and using the standard definition
(i.e. region within $\approx$3 log likelihood units of the likelihood 
maximum). 
Since we are following closely the approach of Yoo04 these should, 
and in fact do,
turn out to be broadly similar to those given in Yoo04. More importantly, 
we show how the joint confidence levels change if we neglect the candidate
which we have shown is a spurious binary. The constraints on MACHOs
are eased substantially, the upper limit on MACHO 
mass has moved out to $\approx 500M_{p}$. 

Figure~\ref{macho_conf} shows that the exclusion of just one binary
has a significant effect on the MACHO constraints. To explore
this effect further, we also examined the impact of removing NLTT
16394/16407 from the sample and find in this case the constraints from
binaries at the 95\% level vanish.  Such sensitivity clearly means
larger samples of wide binaries are urgently needed to solidify the
constraints on MACHOs.

\subsubsection{Orbits: A Note of Caution}
\begin{figure}
\includegraphics[width=0.47\textwidth]{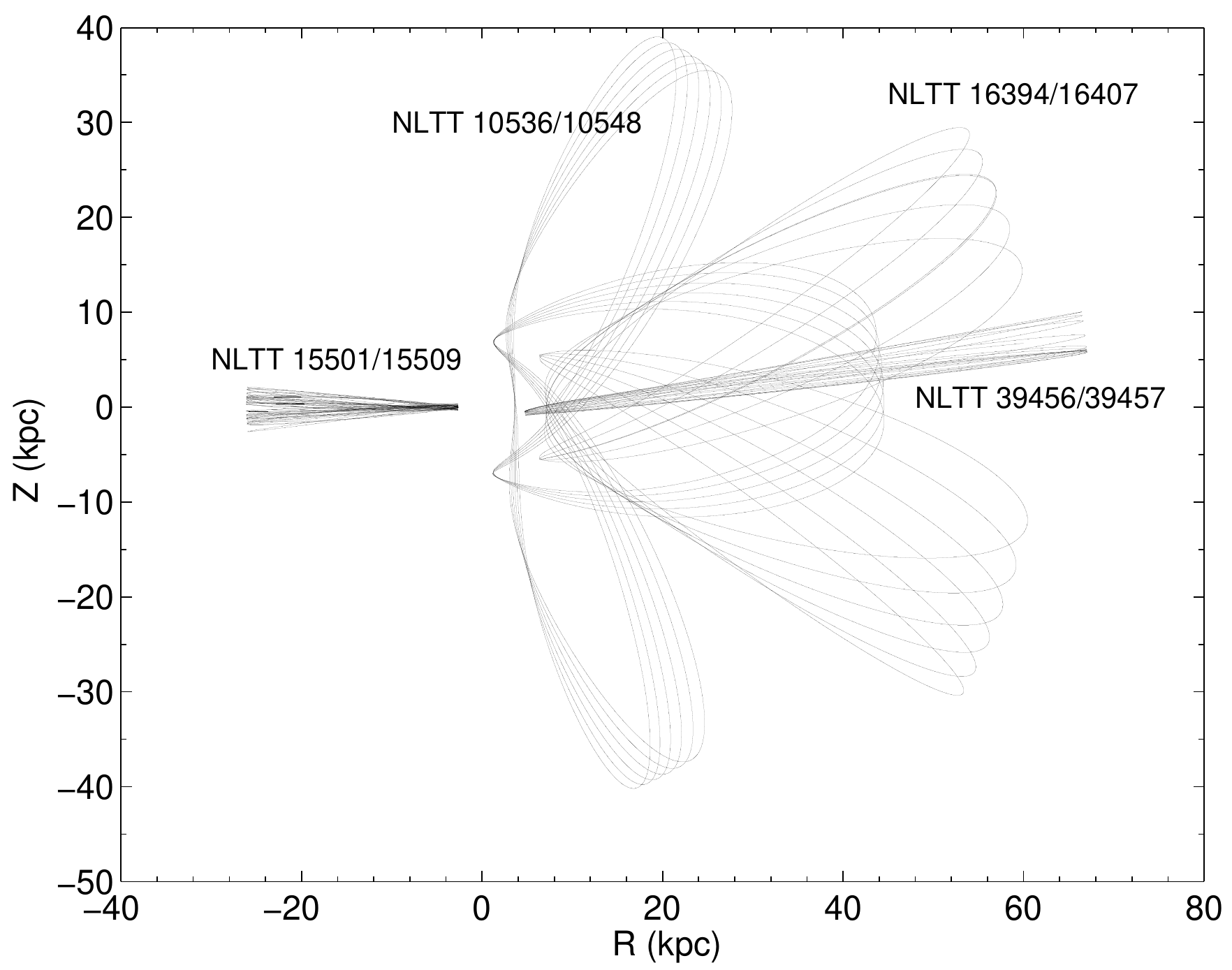}
\caption[]{\label{wide_orbit} Orbits over  10 Gyrs for the 3 
wide binaries that we confirmed and wide binary  NLTT 39456/39457.
The Milky Way Mass model 1 of \cite{DehnenMW} is assumed and for 
clarity we have flipped the sign of R for NLTT 15501/15509.}
\end{figure}

With proper motions, radial velocity and an estimate of the distance
we can plot the orbit for the confirmed binaries
(Figure~\ref{wide_orbit}).  The orbits confirm that the binaries
belong to the halo. For NLTT 16394/16407 we find that along the orbit
the average dark matter density experienced by the object over a 10
Gyr period is 10$\%$ of the local dark matter density.  (Even if we
assume the distance to this binary is 20\% less than predicted by the
CG04 relation the average dark matter density is still only 40\% of
the local dark matter density.) This implies that the inclusion of
this object in the sample and the use of the local solar density are
incompatible. In fact, the two other binary pairs in our sample
experience time-averaged dark matter densities of 45\% and 16\% of the
local density, while for NLTT 39456/39457 it is 11\%. If these orbits
are representative of the orbits of the widest binaries in the sample
then this trend could be a sign that the widest binaries can only
survive by spending most of the orbit away from the inner regions of
the Galaxy. If we take the mean of the time-averaged halo density
experienced by the four binaries as a more representative value for
the dark matter encountered by a typical halo binary along its orbit,
we can still use the constraints discussed above but the contours
defined by the binary constraints plotted in Figure~\ref{macho_conf}
need to be shifted upwards by a factor of five. This would seriously
undermine the constraints that can be drawn from wide
binaries. 

\section{Conclusion}
\label{sec:conc}
A population of MACHOs with masses beyond the current micro-lensing
detection threshold could have a marked effect on the separation
distribution of wide halo binaries. While the actual number of
observed candidate wide halo stellar binaries is small, strong
constraints on MACHOs have been drawn from their distribution. We have
measured the radial velocities of four of the widest candidate wide
halo binaries from the sample used to place the existing constraints.
These measurements provide a consistency test on the binarity of these
objects and provide the data needed to examine their Galactic
orbits. Our data confirm that three of the four widest halo
binary candidates in the CG04 sample are real, thereby vindicating the
search strategy of CG04 and demonstrating explicitly that binaries
with separations of $\gtrsim 1pc$ can exist. However, the spurious
nature of the second-widest pair and the orbit of the widest object
undermines the existing constraints on MACHOs from analysis of wide
halo binaries. The current wide binary sample is too small to place
meaningful constraints on MACHOs; in particular the constraints are
extremely sensitive to the widest binary in the sample which, as we
have shown, experiences a much lower dark matter density than the
value in the analysis leading to the constraints. Increasing the size
of the wide binary sample, for example using the SDSS proper motion
data or, in the longer term, using \textit{Gaia}, is thus essential if
we are to constrain the clumpiness of the dark matter distribution in
the Milky Way and determine whether our results are just a reprieve
for MACHOs.

\section*{Acknowledgments}
Our WHT data were obtained as part of the ING service program and we
gratefully acknowledge the ING staff for taking these data. We thank
J. Yoo and J. Chaname, in particular, for providing 
information about the original CG04 data set. 
We thank L. Wyrzykowski for providing microlensing
data.We also thank the anonymous referee for a prompt and insightful
report.  MIW is supported by the Royal Society.

\end{document}